\pdfoutput=1
\documentclass[twocolumn, showpacs,preprintnumbers, amsmath, epsfig, floatfix, prl]{revtex4}
\usepackage{dcolumn} 
\usepackage{bm}
\usepackage{mhchem}
\usepackage{graphicx} 
\usepackage{epstopdf} 
\usepackage{xcolor} 

\begin{document}

\title{
Electric Field Tuning of the Rashba Effect in the Polar Perovskite  Structures
}
\date{\today}

\author{K. V. Shanavas$^\dagger$ and S. Satpathy}
\affiliation{Department of Physics, University of Missouri, Columbia, MO 65211, USA}

\begin{abstract}
We show that the Rashba effect
at the polar perovskite surfaces and interfaces  can be tuned  by manipulating the  two-dimensional electron gas (2DEG) by an applied electric field, using it to draw the 2DEG out to the surface or push it deeper into the bulk, thereby controlling the surface-sensitive phenomenon.  These ideas are illustrated by a comprehensive density-functional study of the recently-discovered polar KTaO$_3$ surface.    Analytical results obtained with a   tight-binding model  unravel the interplay between the various factors affecting the Rashba effect  such as the strength of the spin-orbit interaction and the surface-induced asymmetry. Our work helps interpret the recent experiments on the KTaO$_3$ surface as well as the SrTiO$_3$/LaAlO$_3$ interface. 
\end{abstract}
\maketitle
%: Introduction
The Rashba effect describes the momentum-dependent spin splitting of the electron states at a surface or interface and 
is the combined result of the  spin-orbit interaction (SOI) and the inversion-symmetry breaking\cite{Rashba}.
It is commonly described by the Hamiltonian  %: Eq. 1
 \begin{equation}
  {\cal H}_R = \alpha_R (\vec k \times  \vec \sigma)    \cdot \hat z,
 \label{Rashba}
 \end{equation}
 where $\vec k$ and $\vec \sigma$ are the electron momentum and spin, $\hat z$ is along the surface normal, and $\alpha_R$ is the Rashba coefficient, which leads to the linear spin splitting in the band structure 
$\varepsilon_k = \frac{\hbar^2 k^2}{2m} \pm \alpha_R k$.
The control of the Rashba effect  by  an applied electric field is at the heart of a class of proposed spintronics devices for manipulating the electron spin\cite{Winkler}. 
The perovskite interfaces\cite{Hwang1, Hwang2} are expected to have a much larger Rashba effect than their semiconductor counterparts\cite{Nitta},  
owing to the presence of high Z elements and a strongly localized 2DEG formed by the polar catastrophe.
In fact, a  strong Rashba effect was recently observed in the \ce{LaAlO3}/\ce{SrTiO3}  interface\cite{LAOSTO-1, LAOSTO-2}, which also showed  an ill-understood asymmetric dependence on the direction of the applied electric field.

In this Letter, we show that the polar perovskite structures constitute an excellent system for the field control of the Rashba effect, aided by the relative ease with which the  2DEG can be manipulated in these polar structures.   Detail density-functional results are presented for the  KTaO$_3$ (KTO) surface to illustrate the ideas. 

%: Fig1
\begin{figure}[h] 
\scalebox{0.6}{\includegraphics{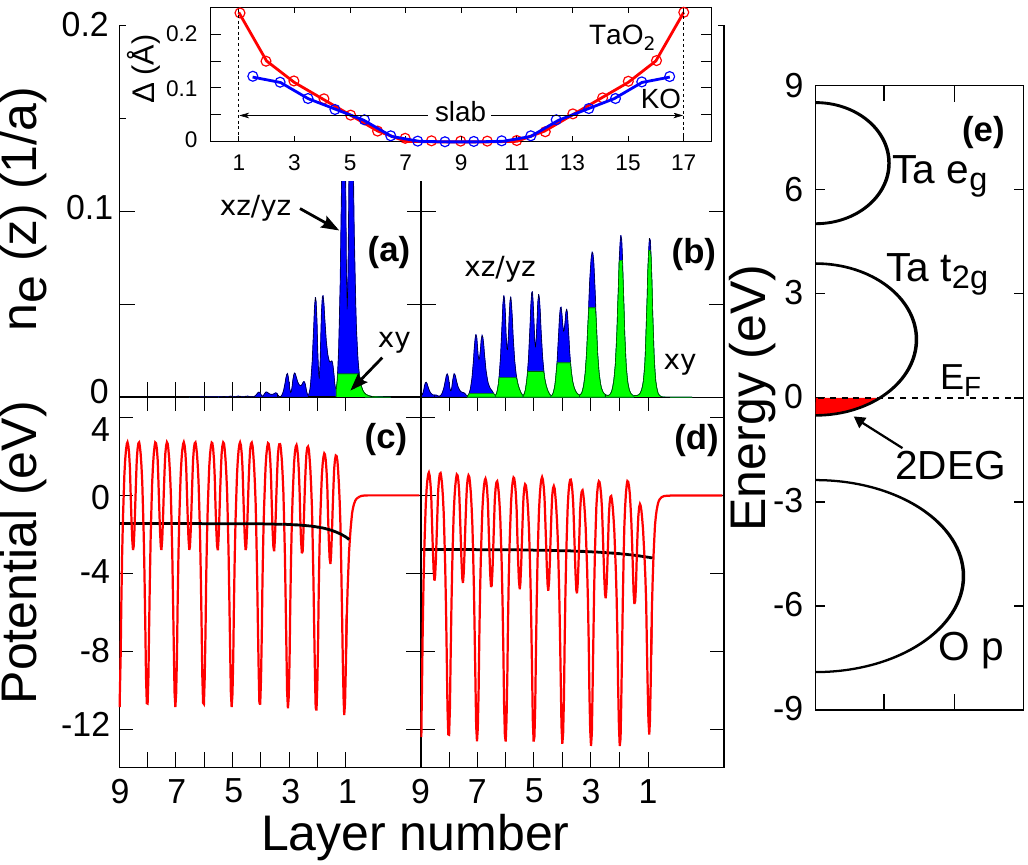}}
\caption{ (Color online) 
Summary of the density-functional results for the KTaO$_3$ surface. The rightmost panel (e) shows the schematic band structure and the formation of the Ta t$_{2g}$-derived 2DEG at the surface. Panels (a) and (c) show the layer density profile of the 2DEG and the surface potential for the unrelaxed  structure, while (b) and (d) show the same for the relaxed case.  The topmost part shows the change of the cation-anion distances $\Delta$ in various layers due to  relaxation, with
layer nos. 1 and 17 being the two surface Ta layers. 
}
\label{densityprofile}
\end{figure}
{\it 2DEG at the KTO surface} -- 
The KTaO$_3$ (KTO) surface  is an ideal system for the study of the Rashba effect   because Ta is a high Z element with strong SOI, a polar-catastrophe induced 2DEG has been observed there recently\cite{King, KTO-PRB} similar to the LAO/STO interface, and finally a surface rather than an interface is more easily amenable to external electric fields. 
Fig. \ref{densityprofile} shows the basic features of the 2DEG formed at the KTO surface obtained from our calculations using density-functional theory (DFT),
 performed with the GGA functional and the projector augmented wave pseudopotential method as implemented in the Vienna ab initio simulation package~\cite{Kresse1993,Kresse1996}. To simulate the \ce{TaO2}-terminated  surface,  we used a slab geometry consisting of 17 \ce{TaO2} and 16 KO alternating layers corresponding to the formula unit (KTO)$_{16.5}$ and 24 \AA\ of vacuum. We studied the Rashba effect by applying a series of electric fields and by fully relaxing the crystal structure in each case.\cite{Supp} 
 
For the KTO surface, the alternating charged layers, nominally (\ce{TaO2})$^{+1}$ and (KO)$^{-1}$,
lead to the polar catastrophe just like in  LAO/STO and as a result a 2DEG forms in the surface region terminated by  \ce{TaO2}. 
Considerable structural relaxation, as expected for a polar surface, spreads the 2DEG several layers into the bulk.
The relaxations, which produce  local  dipole moments screening out the surface polar field,  decay rapidly within about six KTO layers 
and beyond that, the ionic positions return to their bulk values.
%: Fig2
\begin{figure}[tb] 
\scalebox{0.45}{\includegraphics{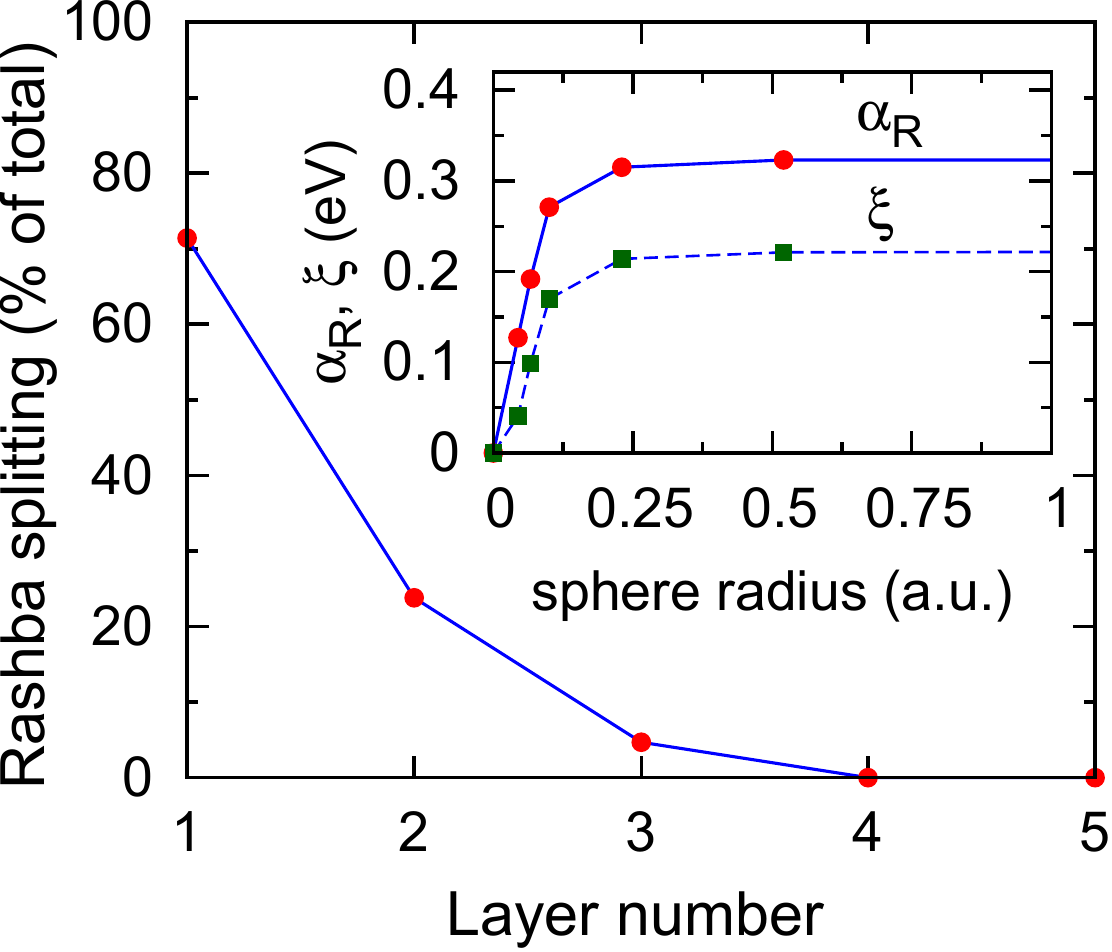}}
\caption{(Color online) Contribution from the various surface layers to the Rashba splitting of the lowest band in Fig. \ref{bands} (a). Inset shows the Rashba coefficient $\alpha_R$ as well as the SOI parameter $\xi$  as a function of the Ta sphere radius within which the nuclear electric field term
$r^{-1} \partial V/ \partial r$  was retained.    
}
\label{PartialRashba}
\end{figure}

%: Rashba effect
{\it The Rashba effect} -- 
The microscopic origin of the Rashba effect is the relativistic SOI 
$
 {\cal H}_{SO}=\frac{\hbar^2}{2m^2c^2} (\vec\nabla V \times \vec {k}) \cdot \vec \sigma,
$
where $\vec \nabla V$ is the potential gradient. 
For a spherically symmetric potential, such as the field from a nucleus, it assumes the familiar form:
${\cal H}_{SO}=(m^2c^2 r)^{-1}  (\partial V /\partial r)  \vec L \cdot \vec S = \xi \ \vec L \cdot \vec S.  $ 
In the presence of a symmetry-breaking surface electric field $E \hat z$, the  first expression for $ {\cal H}_{SO}$ leads to  Eq. (\ref{Rashba}),
 with  the Rashba coefficient $\alpha_R = - \frac{\hbar^2 E}{2m^2c^2}$.
However, this coefficient is severely underestimated, if one naively identifies the electric field with the surface potential gradient.

\begin{figure}[bt] 
\scalebox{0.48}{\includegraphics{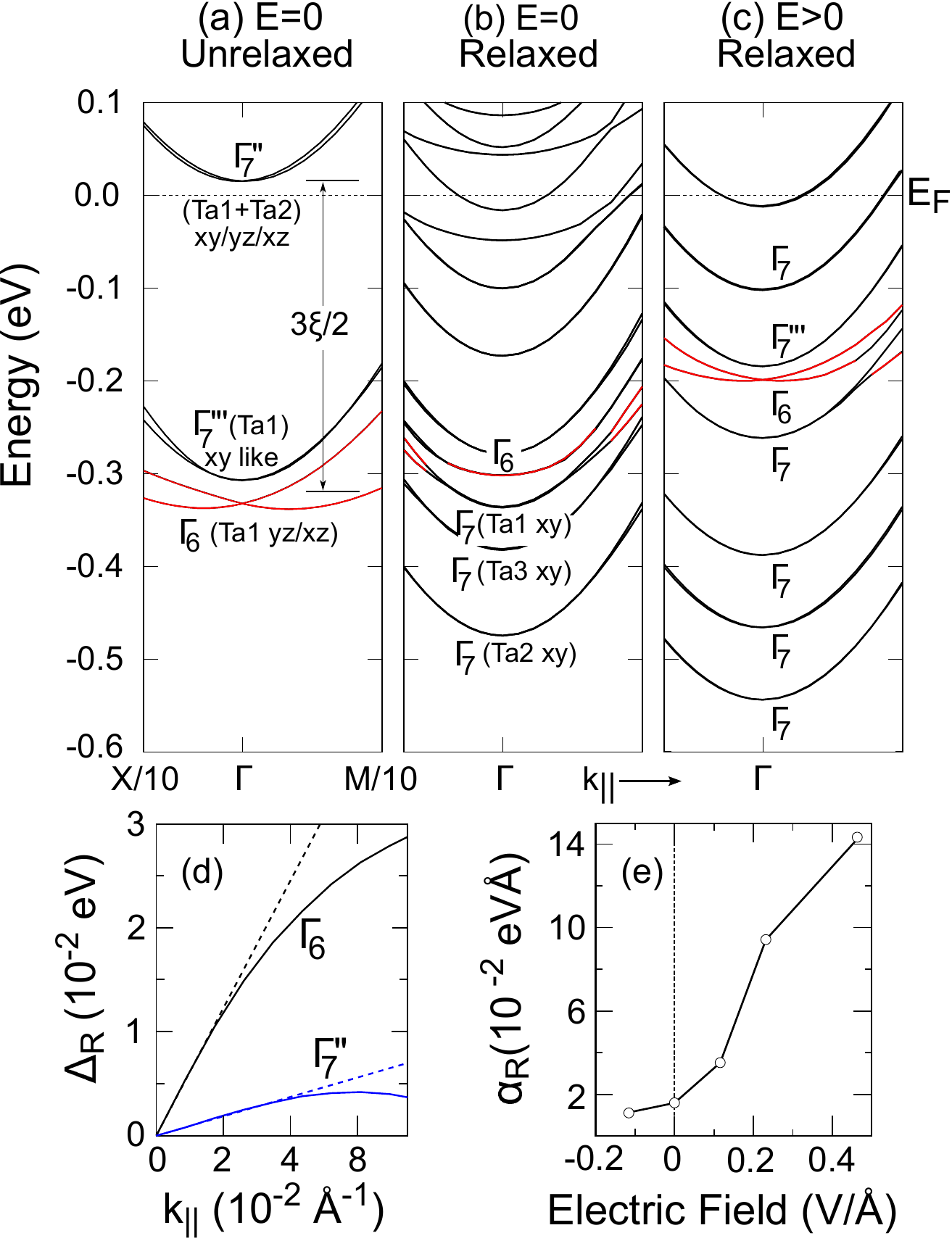}} 
\caption{(Color online) Effect of surface relaxation and the applied electric field on the Rashba splitting for the KTO surface as obtained from DFT. Bands with strong Rashba splitting are shown in red. Relaxation causes the 2DEG to migrate deeper into the bulk  diminishing the Rashba splitting (b), while an applied electric field
($E = 0.5$ V/\AA) draws it back to the surface enhancing the splitting
 (c). Fig. (d) shows the splitting $\Delta_R$ as a function of $k_{||}$ for bands in (a), the slopes of which yield $\alpha_R = 0.3$ eV.\AA\ for $\Gamma_6$ and 0.05 eV.\AA\ for $\Gamma_7''$. Fig. (e) shows the change of $\alpha_R$ with the applied electric field (positive $E$ points into the bulk). The $k$ points  correspond to $X = (1,0)$ and $M = (1,1)$ in units of $2\pi a^{-1} = 2.56$ \AA$^{-1}$. 
}
\label{bands} 
\end{figure}

Rather, the correct picture is that the Rashba SOI originates in the  nuclear region due to the large nuclear field gradient there\cite{Blugel}.
The applied electric field  polarizes the orbital wave functions, changing their weights at the nucleus, so that the electron experiences a different nuclear field. We illustrate this in Fig. \ref{PartialRashba}  by computing  the various contributions to the Rashba splitting  for the  $\Gamma_6$ bands in Fig. \ref{bands} (a).  We have isolated  these contributions by keeping the SOI $\xi$ either (i) on atoms in specific layers or (ii) on all atoms  but within a specified spherical nuclear region  and then by performing a single iteration with the self-consistent DFT potential obtained with all interactions present.  As Fig. \ref{PartialRashba} shows, the dominant contribution comes from the nuclear region of the surface atoms. This in turn suggests the tuning of the Rashba effect by an electric field by moving the 2DEG in and out of the surface layers.       
 
%: Fig3

%: Fig4
\begin{figure} 
\scalebox{0.65}{\includegraphics{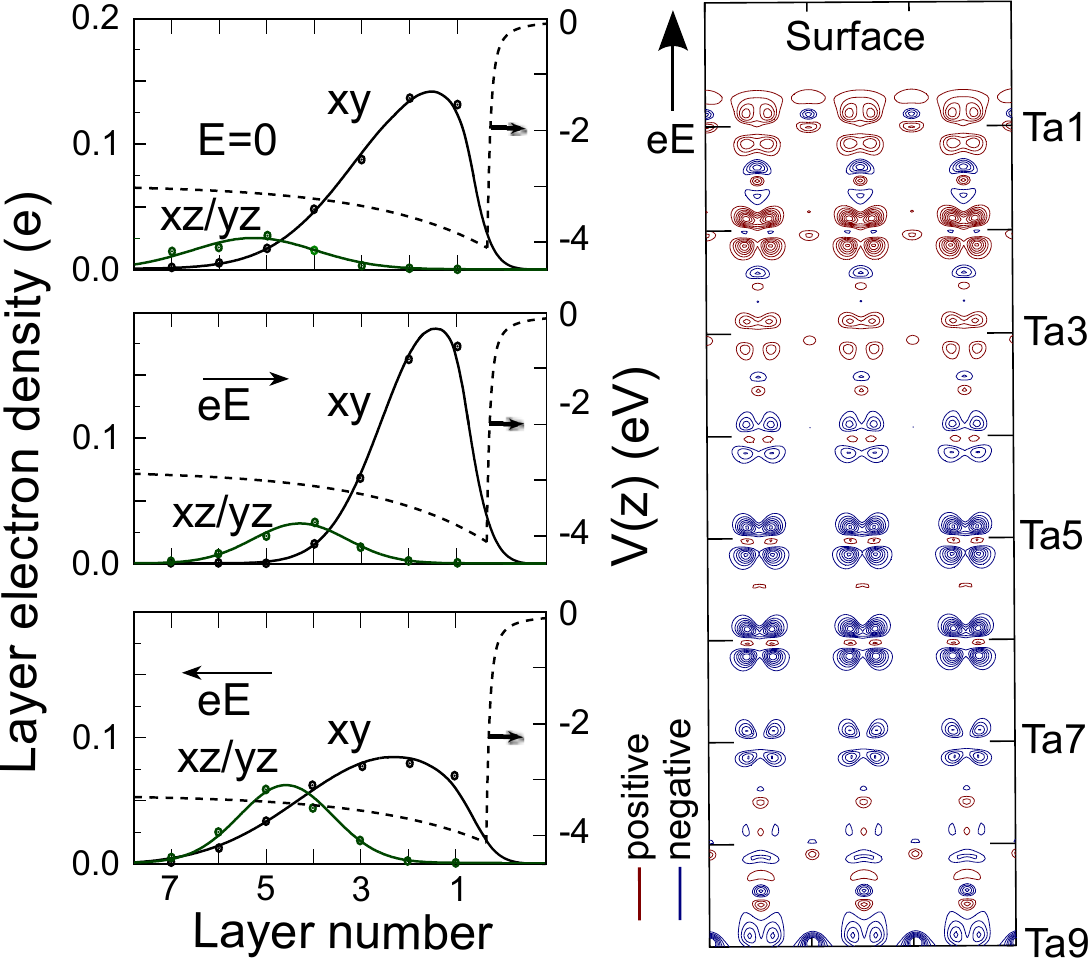}}  
\caption{ (Color online)   
{\it Left}  shows the layer density profile of the 2DEG (solid dots) with and without an applied electric field ($E = 0.12$ V/\AA) calculated from DFT. 
The dashed lines indicate the cell-averaged potentials, while  the solid lines are guides to the eye, with  the black lines also indicating the electron leakage out of the surface obtained from solving the 1D Schr\"odinger equation with the surface potential. {\it Right} shows  contours of the electron density change   due to the applied electric field, which  drives the electrons   to the surface.
}
\label{Contours}  
\end{figure}
%
%:  Electric field tuning 
%
{\it Electric field tuning} -- 
We have calculated the Rashba splitting for the KTO surface by applying a series of electric fields.
As seen from Fig. (\ref{bands}), the unrelaxed surface with zero field shows  a very strong linear-$k$ Rashba splitting because the 2DEG is sharply localized at the surface due to the strong polar field, extending to just three TaO$_2$ layers (Fig. \ref{densityprofile}). Relaxation of the surface atoms screens out the polar field and as a result, the 2DEG spreads deeper into the bulk region, thereby washing away the Rashba effect. This explains why in the ARPES experiments\cite{King,KTO-PRB} on KTO, the Rashba splitting has not been seen despite the presence of a large spin-orbit coupling. On the contrary, application of an electric field draws the 2DEG towards the surface (Fig. \ref{Contours}), 
restoring back the Rashba effect. An electric field in the opposite direction drives the 2DEG deeper into the bulk and the Rashba splitting quickly becomes very small.
Thus, we have demonstrated the field tuning of the Rashba effect as well as the very interesting asymmetric dependence on the direction of the applied electric field (Fig. \ref{bands} (e)).  Such an asymmetric dependence  was recently observed in the LAO/STO interface\cite{LAOSTO-2}. Note that the asymmetry is not expected for a non-polar surface such as Ag and a symmetric Rashba effect has been predicted there.\cite{Ag} 
%: Tight-binding description
{\it Tight-binding description} --
The Rashba splitting differs widely within the $d$ orbital manifold, which may be understood in terms of the tight-binding (TB) model\cite{Petersen} on the cubic lattice that includes the surface asymmetry and the electric field:
\begin{equation}
 {\cal H} =  {\cal H}_{ke} +  {\cal H}_{SO} +  {\cal H}_E + V_{sf}.
\label{HTB}
\end{equation}
The kinetic energy part 
contains the standard $V_\sigma$ and $V_\pi$  hopping between the $d$ orbitals
and the crystal field energies: ${\cal H}_{ke} = \sum_{ip\sigma} \varepsilon_{ip} n_{ip\sigma} + $ $ \sum_{ip,jq;\sigma} V_{ij}^{pq}  c^\dagger_{ip\sigma}c_{jq\sigma} + h. c.$,  where $ip\sigma$ denotes the site-orbital-spin index. The $O_h$ cubic  field splits the d states into  $e_g$ plus $t_{2g}$ states. With the SOI included, the six-fold degenerate $t_{2g}$ states (including spin) split into a two-fold $\Gamma_7^+$  and a four-fold $\Gamma_8^+$ state, while the e$_g$  remains unsplit with $\Gamma_8^+$  symmetry.  The surface reduces the cubic symmetry into $C_{4v}$ with $\Gamma_7^+$ going into $\Gamma_7$,  while the $\Gamma_8^+$ state splits into $\Gamma_6 + \Gamma_7$, both two-fold degenerate.\cite{Koster} Note that there are just two double representations  $\Gamma_6$ and $\Gamma_7$ for the $C_{4v}$ group; we have used primes on $\Gamma_7$ to indicate its different orbital character (see Table I) due to the symmetry-allowed mixing between the two 
 $\Gamma_7$ states in the $t_{2g}$ manifold.\cite{Supp}

 %=========================================================================== 
%: Table I
\begin{table} [t]
\caption{Rashba coefficient $\alpha_R$ and the pseudo-spin partner functions for the $d$ states.
Energies of the spin-orbit split states appear in the parenthesis  and $a$ is the lattice constant.  
If the SOI $\xi$ is strong, but the electric field (parametrized by $\alpha, \beta, \gamma$) is {\it not} weak, the 
$\Gamma_7^{'''}$ and the $\Gamma_6$ states do not reduce to the Rashba form, but must be described by a $4 \times 4$ matrix (Eq. \ref{44}),  % in the joint $\Gamma_7^{'''} \otimes \Gamma_6$ subspace
while the $\Gamma_7^{''}$ has the same $\alpha_R$ as in Case 2. Strong cubic field splitting $\Delta \gg \xi$ is assumed.
}
\begin{center}
\begin{tabular}{ ll | ll | c}
\hline 

\multicolumn{2}{c|}{Cubic Field} &\multicolumn{2}{c|}{Surface Field  
$(C_{4v})$} & Rashba   \\
\multicolumn{2}{c|} { $(O_h)$ }&   & Rashba pseudo-spin   &  coefficients  \\  
\multicolumn{2}{c|}{  }& Symmetry & partner functions   & $\alpha_R/a$   \\  
%\multicolumn{2}{c|} {  }& &  & \\
\hline
\hline
   $e_g$ & 
$\Gamma_8^+ (\Delta)$ &$\Gamma_7 (\Delta + \delta)$ &     $ z^2\uparrow, z^2\downarrow$ &   $ -2\sqrt3\beta  \xi/\Delta$ \\ 
&
& $\Gamma_6 (\Delta)$ &     $x^2-y^2\uparrow, x^2-y^2\downarrow$ &   
$-2\gamma \xi/\Delta$    \\ \hline

\multicolumn{4}{l|}{$t_{2g}$ Case 1.   Weak SOI, $ \xi \ll |\varepsilon| $}
 \\
 
$\Gamma_7^+$ &   $(\xi)$ \phantom{----}
& $\Gamma_7 (-\varepsilon)$ &  $xy\uparrow, xy\downarrow$ &   $ 2\alpha  \xi/\varepsilon   $    \\

 $\Gamma_8^+$ & $(-\xi/2)$  &    $\Gamma_7^\prime (\xi/2)$ &  \hspace {-5 mm} $      yz\downarrow +  ixz\downarrow, yz\uparrow -ixz\uparrow$ &   
$  2\alpha  \xi/\varepsilon$    \\

&  & $\Gamma_6 (-\xi/2)$ &  \hspace {-5 mm}  $ yz\downarrow -
ixz\downarrow, yz\uparrow +ixz\uparrow $ &  
 $ 2\sqrt3\beta  \xi/\Delta$       \\
\hline
 
\multicolumn{5} {l}{ $t_{2g}$ Case 2. Strong SOI, $ \xi \gg |\varepsilon| $, weak electric  field  $ |\alpha| \ll |\varepsilon |$ }   \\ 

$\Gamma_7^+$ &   $(\xi)$ \phantom{----} & $\Gamma_7^{''}( \xi -\varepsilon/3 )$  & 
$xy\uparrow + yz\downarrow + i xz \downarrow$, &  $-4\alpha  /3   $    \\

 &  & & $xy\downarrow  - yz\uparrow + i xz \uparrow$ &  \\

$\Gamma_8^+$ &  $(- \xi/2)$  &  $\Gamma_7^{'''}  (-\frac{ 3\xi + 4\varepsilon} {6})  
   $ &  $2xy\uparrow -yz\downarrow   -ixz\downarrow$,
&   $4\alpha  /3   $     \\

  &  &  & $2xy\downarrow +yz\uparrow -ixz\uparrow$   &  \\
  &  &  $\Gamma_6 (-\xi/2)$ &    \hspace {-5 mm}   $ yz\downarrow -
ixz\downarrow, yz\uparrow +ixz\uparrow$ &   $ 2\sqrt3\beta  \xi/\Delta$       \\

 \hline

\end{tabular}
\end{center}
\end{table}
%: Table I End

%: Fig5 
\begin{figure}[b] 
\scalebox{0.55}{\includegraphics{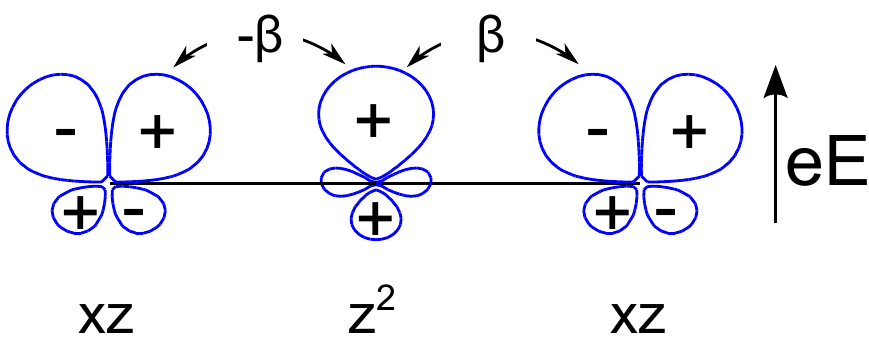}}
\caption{ (Color online) Electric-field-induced hopping between the $d$ orbitals due to  the orbital polarization. 
}
\label{hopping} 
\end{figure} 

The remaining  parts, $V_{sf}$ and ${\cal H}_E$,  in Eq. (\ref{HTB})  are the inversion symmetry breaking fields crucial for the Rashba effect.
The surface field $V_{sf}$ is modeled by  an asymmetric energy for the surface orbitals: $\varepsilon =  \varepsilon (xz/yz) - \varepsilon (xy)    $  and  $\delta = \varepsilon (z^2) - \varepsilon (x^2-y^2) $, an asymmetry that may come from  strain, the electric field via the atomic relaxation it produces, or  the hopping differential between the orbitals, e.g., $xy$ and $xz/yz$ \cite{Popovic2008} and as such has a complex dependence on the electric field. The electric field part ${\cal H}_E$ induces new hoppings (Fig. \ref{hopping}) between  atoms: $\alpha=\langle xy|{\cal H}_E|xz\rangle_{\hat y}$, $\beta=\langle xz|{\cal H}_E|z^2\rangle_{\hat x}$, and $\gamma=\langle x^2-y^2|{\cal H}_E|yz\rangle_{\hat y}$, whose strengths are roughly proportional to the local electric field. Here  the subscript denotes the direction of the nearest-neighbor on which the second orbital is located. Typical parameters for KTO are\cite{Neumann, Jellison}: $\Delta = \varepsilon (e_g) - \varepsilon (t_{2g} )\approx 4$ eV, $\xi \approx 0.26$ eV, $V_\sigma \approx -1$ eV,  $V_\pi \approx -0.5$ eV,  while  $\alpha, \beta,$ and $ \gamma $ are $\sim 10$ meV at the surface layer.
As one goes into the bulk, the surface asymmetry parameters $\varepsilon, \delta, \alpha, \beta,$ and $\gamma$ rapidly go to zero, so that the Rashba effect comes just from the first few surface layers. 
 % 
 
\begin{figure}[tb] 
\scalebox{0.5}{\includegraphics{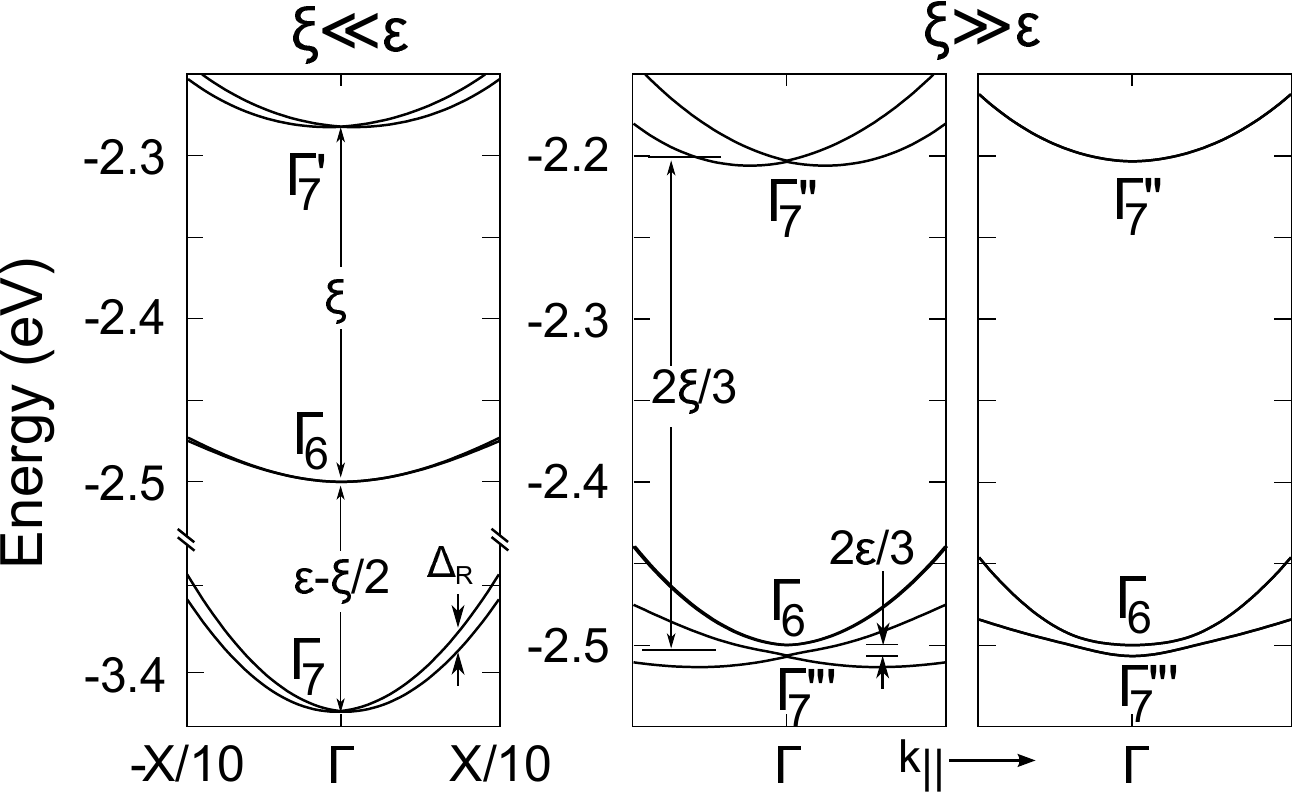}}
\caption{ Rashba splitting of the t$_{2g}$ states in two different regimes of the SOI strength $\xi$. Fig. (a) corresponds to a weak  $\xi \ll \varepsilon$ (in eVs, $\xi = 0.2, \alpha = 0.05, \varepsilon =1$) (Case 1 in Table I).
In (b) and (c), since $\xi$ is strong  ($\xi \gg \varepsilon$), but the electric field, parametrized by $\alpha$, is  not weak ($\xi = 0.2, \alpha = 0.05, \varepsilon =0.01$),  the splitting of the  near-degenerate four-fold bands must be described by 
Eq. \ref{44}. Fig. (c) has the same parameters as (b) except  $\alpha =0$, so that there is no Rashba splitting.
}
\label{TB-44}
\end{figure}

%:Lowdin
We obtain the Rashba  splitting from  Eq. \ref{HTB} by L\"owdin downfolding\cite{Supp, Lowdin} of the effects of the higher-energy bands. The results can be expressed in the Rashba form $  {\cal H}_R = \alpha_R (\vec k \times  \vec \sigma)    \cdot \hat z$  for most bands and the corresponding  Rashba coefficients and the  partner functions for the pseudo spin $\vec \sigma$ are listed in Table I. However, for  near-degenerate cases,  where the SOI is strong ($\xi \gg |\varepsilon|$) but the surface field does not sufficiently lift the four-fold degeneracy of the $\Gamma_8^+$ state ($|\varepsilon| \ll |\alpha|$ or $|\varepsilon| \sim  |\alpha|$), the L\"owdin downfolding  fails and the Rashba SOI can only be written as a $4 \times 4$ matrix spanning the $\Gamma_8^+$ subspace:
%: 4 x 4 matrix
\begin{equation}
 {\cal H}= \frac{2}{ 3}
\begin{pmatrix}
ak^2 + \varepsilon & 2\alpha\,k_+ & c\bar k^2 & - \sqrt 3 \alpha\,k_+      \cr 
2\alpha\,k_- & ak^2 + \varepsilon & \sqrt 3 \alpha\,k_+ & -c\bar k^2 \cr 
 c\bar k^2 & \sqrt {3} \alpha k_- & bk^2 & \frac{3\sqrt 3 \beta \xi }{\Delta} k_+ \cr
  - \sqrt 3 \alpha\,k_+ & -c\bar k^2 & \frac{3\sqrt 3 \beta \xi }{\Delta} k_-   &  bk^2
 \end{pmatrix}.
 \label{44}
\end{equation}
Here, we have included the quadratic-k terms (the band mass), 
with $k_\pm = k_y \pm i k_x$, $\bar k^2 = (k_+^2 - k_-^2)$,  $a = -5V_\pi /3$, $b= -V_\pi /2$, and $c=  - \sqrt{3} V_\pi/4 $, and the order of the basis is the same as the order of appearance of the four $\Gamma_8^+$ partner functions in Case 2, Table I. The Rashba part ${\cal H}_R$ is simply   Eq. (\ref{44}) minus the $k^2$ terms. Eq. (\ref{44}) is valid for strong $\xi$ and for any  $\varepsilon $ and $ \alpha$. If $|\varepsilon| \gg |\alpha|$, one recovers  the results of Case 2, Table I using  L\"owdin downfolding. Fig. \ref{TB-44} shows the TB bands for cases relevant to the Rashba splitting seen in the DFT bands.

%: Fig6
Note from Table I  that even though the linear-k Rashba splitting is always present, its magnitude is very small ($\sim 1/\Delta$) for the $e_g$ bands as well as for the $t_{2g}$-derived $\Gamma_6$ bands. For these bands, the higher-order $k^3$ term may  in fact be dominant as has been seen in the SrTiO$_3$ surface\cite{STO-k3}   
and also suggested by Zhong et al\cite{Zhong}. Also as Table I shows, the pseudo-spin partner functions are sometimes not spin entangled at the $\Gamma$ point, but they always become so away from $\Gamma$ due to the spin mixing via the Rashba Hamiltonian. Returning to the $t_{2g}$ bands, which make up the 2DEG,  for small SOI $\xi$ relative to the surface field $\varepsilon$ (Case 1 in the Table), the Rashba coefficient can be small if $\varepsilon$ is large (note that $\varepsilon =  \varepsilon (xz/yz) - \varepsilon (xy) $  can be varied widely in a material due to lattice relaxation, electric field, or strain, while $\xi$ is more or less fixed). This is the reason for the relatively weak Rashba splitting seen in the TB bands (Fig. \ref{TB-44}) and also in the DFT bands for the relaxed case with $E=0$ (Fig. \ref{bands} (a)). 
The Rashba effect is enhanced significantly in the opposite limit ($\xi \gg |\varepsilon|$), if at the same time the surface field $\varepsilon$  is small or comparable to the electric-field-induced hopping $\alpha$. In this scenario, the Rashba effect is described by Eq. (\ref{44}). 
This is the case for  Fig. \ref{TB-44} (b) and (c), where a large Rashba splitting is seen for the $\Gamma_7'''$ bands, and also for the DFT bands (Fig. \ref{bands} (a) and (c)). Thus the electric field changes the Rashba effect in two ways: {\it one}, by changing the density of the 2DEG  in the surface layers and {\it two}, by altering the surface asymmetry field  $\varepsilon$ and reorienting the orbital energies.
%: Conclusion
 In conclusion, we showed that the Rashba effect
can be tuned in the polar perovskite oxides by manipulating the 2DEG profile by an external electric field.   
These results are relevant not just for the KTO surface, but also for polar materials in general that contain surface or interface $d$ electrons. We also note that since the energies of the  $d$ orbitals are sensitive to the applied strain,
this suggests another means of tailoring the Rashba effect.

We thank Zoran Popovic for stimulating discussions and computational help.
This work was supported by the Office of Basic Energy Sciences of the U. S. Department of Energy  through Grant No. 
DE-FG02-00ER45818.

%:References

\newpage

\begin{widetext}
\subsection{Electric Field Tuning of the Rashba Effect in the Polar Perovskite Structures - Supplementary Information}
\end{widetext}

\section{Density-Functional Methods}
The electronic structure calculations presented in this paper were performed using ab initio density functional theory (DFT) as implemented in the Vienna ab initio simulation package (VASP)\cite{Kresse}. This method uses a plane wave basis set along with the   projector augmented waves (PAW)\cite{PAW} in the ionic core region. We used the generalized gradient approximation with the Perdue-Burke-Ernzerhof (PBE) parametrization~\cite{Perdew-PBE} for the exchange-correlation functional.  The optimum values of the energy cutoff and the size of the k-point mesh were found to be 450 eV and $11\times11\times1$, respectively, and were accordingly employed in our calculation. The calculations of the Rashba coefficients were carried out by taking into account the spin-orbit coupling term perturbatively on top of the fully optimized charge densities.

\begin{figure}[h] 
\scalebox{1.2}{\includegraphics{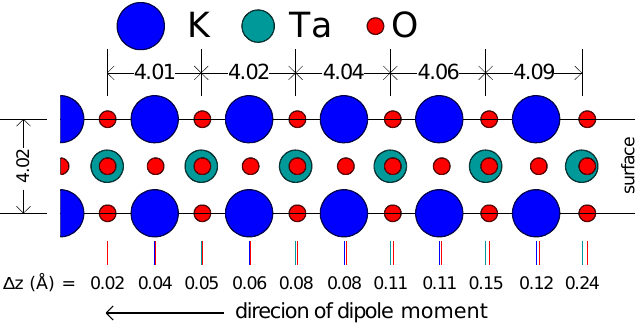}}
\caption{(Color online) Relaxed structure of the KTaO$_3$ slab in the first six layers near the surface, beyond which there was no significant distortion from the unrelaxed structure.} 
\label{str}
\end{figure}

The slab geometry used in the Rashba calculations was created by first optimizing the bulk KTaO$_3$ structure and the cell parameters. Further relaxations of the slab was carried out by keeping the cell dimensions fixed, but allowing all atoms to relax according to the  Hellman-Feynman forces on each atom with the tolerance value of $10^{-2}$ eV\AA$^{-1}$. The atomic displacements are significant only near the first few surface layers as seen from Fig. \ref{str}, where we have shown the  cation-anion displacements for the KO and TaO$_2$ planes in the first six layers near the surface. They are close to zero in the middle five layers in the simulation cell and are not shown here.

Both surfaces of the slab are TaO$_2$ terminated and for the calculations with the electric fields, a V-shaped potential (Fig. \ref{V}) was applied. The linear muffin tin orbitals (LMTO) method\cite{Andersen} was used along with the same exchange-correlation functional as above to obtain the partial contributions to the Rashba coefficients shown in Fig. 2 of the main text. To obtain these partial contributions, the spin-orbit interaction was retained within a specified radius of the nuclei or on atoms in specific layers.

\begin{figure}[h] 
\scalebox{0.4}{\includegraphics{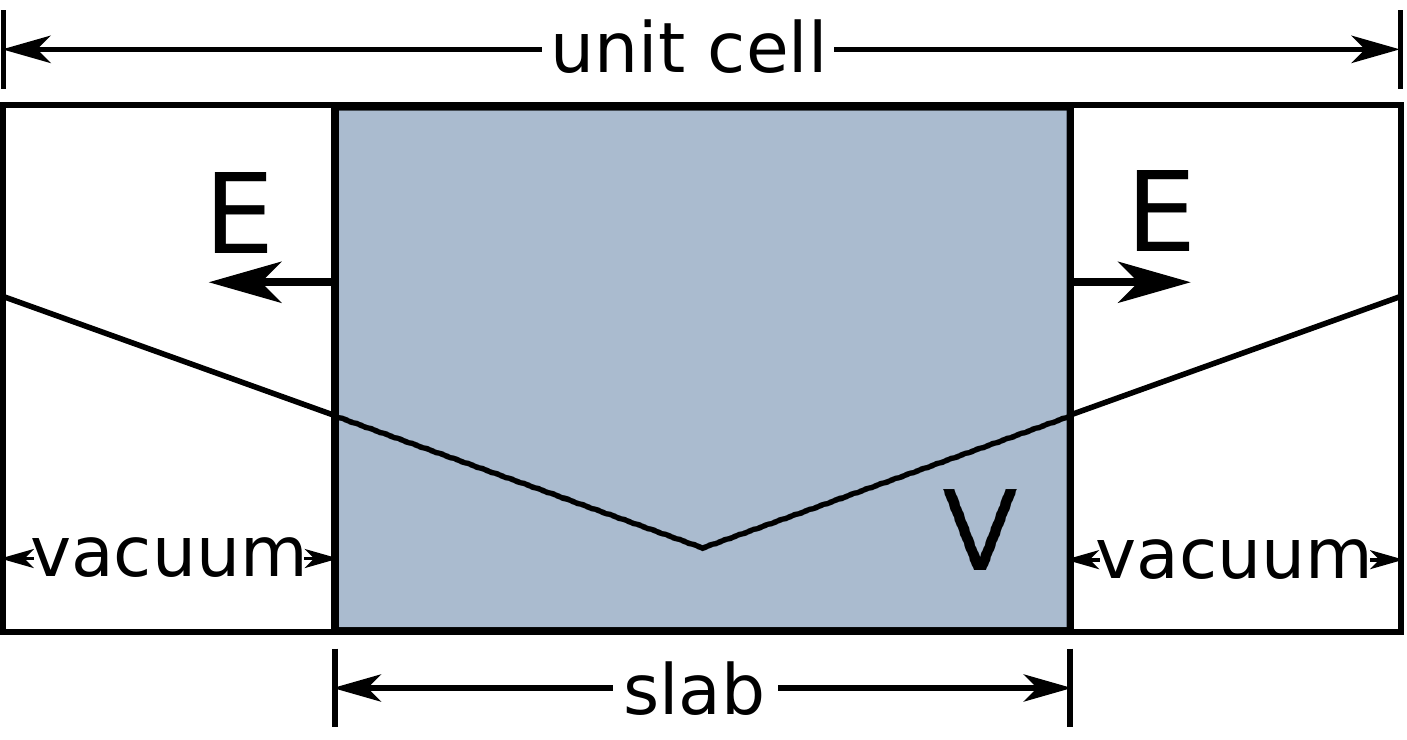}}
\caption{The symmetric V-shaped potential applied to the slab to compute the Rashba coefficients in the density-functional studies.
The slab consisted of sixteen full layers of KTO plus an extra layer of TaO$_2$, so that both surfaces are TaO$_2$ terminated, plus 24 \AA\ of vacuum.}
\label{V}
\end{figure}

\section{Surface potential}
As emphasized already in the main text, the contribution of the surface electric field to the Rashba coefficient is negligible as compared to what is observed in the experiments or computed from the density-functional calculations. Much of the contribution comes from the nuclear regions of the heavy atoms near the surface, where  the inversion symmetry is broken. 

To assess the effect of the surface field, the surface potential was calculated from the density-functional theory. We computed the surface potential by averaging the Kohn-Sham potential over the $xy$-plane parallel to the surface. This planar-averaged potential, shown by the red lines in Fig. 1 in the main text, has strong oscillations due to the presence of the atomic planes. Averaging it further over a unit cell length along the $z$ direction, we get the average surface potential $V(z)$,
which may be fitted to  a modified Jones-Jennings-Jepsen 
form\cite{Jepsen}:
\begin{equation}
V(z)=\left\{ \begin{array}{ll}
[2(z-z_0)]^{-1}   [    -1+e^{-\lambda(z-z_0)} ]       &           z>z_0 \\   
U -   V e^{\kappa(z-z_0)} & z<z_0.
\end{array} \right.
\label{V}
\end{equation}
The potential
has the image form in the vacuum region $ z > z_0$, with $z_0$ being the image plane location, while inside the material, we have the Thomas-Fermi exponential form for metallic screening with the inverse screening length $\kappa$.
For the relaxed KTO  in the absence of an external field, the parameters  are: $U = -3.1$ eV, $V = 1.1$ eV, $z_0 = 1.3$  \AA, $\lambda = 2.1$ \AA$^{-1}$, and $\kappa = 0.10$  \AA$^{-1}$. The screening is much weaker than the Thomas-Fermi uniform electron gas
result $\kappa_0 = 2.95 (r_s/a_0)^ {-1/2}$,  $r_s \sim 7 $ \AA, since the electron gas is present only in a small region near the surface.  
The exponential decay of the electrons into the vacuum region shown in Fig. 1 in the main text  
was obtained 
by solving the 1D  Schr\"odinger equation   $[-\nabla^2+V(z)]\psi=E\psi$ with this potential.

\section{Crystal Field Symmetry}

Under the $O_h$ crystal field, the five $d$ orbitals split into the two-fold $\Gamma_{12}$  ($z^2$, $x^2-y^2$) and the three-fold $\Gamma_{25}^\prime$ ($xy, yz, zx$) states (also referred to as $e_g$ and $t_{2g}$, respectively).\cite{Satpathy} When the spin-orbit coupling term ${\cal H}_{SO}=  \xi \ \vec L \cdot \vec S  $ is included, the $t_{2g}$ states (six states including spin degeneracy) break into a spin-orbit coupled set of states, viz., a two-fold $\Gamma_7^+$ and a four-fold $\Gamma_8^+$,\cite{Satpathy, Koster} with the partner functions as follows:

\begin{align}
\Gamma_7^+ =
\begin{cases}
\left.\begin{aligned} 
\frac{1}{\sqrt 3} (xy \uparrow + yz \downarrow + i xz \downarrow)   \\
\frac{1}{\sqrt 3} (xy \downarrow - yz \uparrow + i xz \uparrow)  
\end{aligned}\right\rbrace \Gamma_7\\
\end{cases}
\label{two}
\end{align}

\begin{align}
\Gamma_8^+ =
\begin{cases}
\left.\begin{aligned} \frac{1}{\sqrt 2} ( yz \uparrow + i xz \uparrow)  \\
\frac{1}{\sqrt 2} ( yz \downarrow - i xz \downarrow) \end{aligned}\right\rbrace \Gamma_6  \\
\left. \begin{aligned} \frac{1}{\sqrt 6} (2 xy \uparrow -  yz \downarrow - i xz \downarrow) \\
\frac{1}{\sqrt 6} (2 xy \downarrow +  yz \uparrow - i xz \uparrow) \end{aligned}\right\rbrace \Gamma_7,   \\
\end{cases}
\label{three}
\end{align}
while the $e_g$ states remain four-fold degenerate with the corresponding partner functions:  
\begin{align}
\Gamma_8^+ =
\begin{cases}
  \left. \begin{aligned}  (x^2-y^2) \uparrow \\
  (x^2-y^2) \downarrow \end{aligned}\right\rbrace \Gamma_6     \\
  \left. \begin{aligned} z^2 \uparrow \\
  z^2 \downarrow \end{aligned} \right\rbrace \Gamma_7.    \\ 
\end{cases}
\label{four}
\end{align}
\begin{figure}[h] 
\scalebox{1.8}{\includegraphics{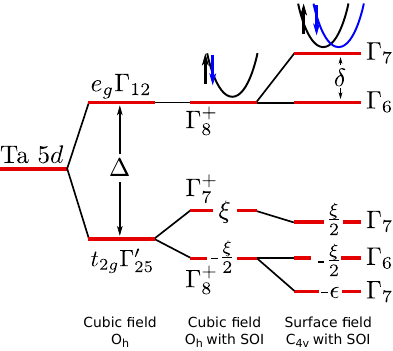}}
\caption{Splitting of the Ta(d) states at the $\Gamma$ point in the Brillouin zone by the crystal field and the SOI. Rashba splitting, indicated by the spin-split parabolas for the $\Gamma_7$ state, occurs when both the SOI as well as broken inversion symmetry are present.   }  
\label{Levels}
\end{figure} 
When the symmetry is reduced from $O_h$ to $C_{4v}$ corresponding to a surface, we have: $\Gamma_8^+ \rightarrow \Gamma_6 + \Gamma_7$
and $\Gamma_7^+ \rightarrow \ \Gamma_7$, with the corresponding partner functions for the $C_{4v}$ group indicated on the right hand side of Eqs. (\ref{two}) - (\ref{four}). Thus, we have  two $\Gamma_7$ states within the $t_{2g}$ manifold, which are allowed to mix by symmetry.  The amount of mixing depends on their relative energies, which are determined by the crystal-field energy $\varepsilon$ and the spin-orbit coupling strength $\xi$. For convenience of discussion, we have used the nomenclatures $\Gamma_7$, $\Gamma_7^\prime$, $\Gamma_7^{\prime\prime}$, and $\Gamma_7^{\prime\prime\prime}$ to indicate different mixtures between these $t_{2g}$-derived $\Gamma_7$ states in Table I of the main text, where also the corresponding partner functions are listed without the normalization factors. (The $e_g$-derived $\Gamma_7$ state does not mix significantly owing to the large energy denominator $\Delta$.) For instance, in Table I, the partner functions for $\Gamma_7^\prime$ are:
\begin{align}
\Gamma_7^\prime =
\begin{cases}
\frac{1}{\sqrt 2} ( yz \downarrow + i xz \downarrow)  \\
\frac{1}{\sqrt 2} ( yz \uparrow - i xz \uparrow),  \\
\end{cases}
\end{align}
while the corresponding orthogonal partners spanning the $t_{2g}$  space are
\begin{align}
\Gamma_7 =
\begin{cases}
 xy \uparrow  \\
xy \downarrow.  \\
\end{cases}
\end{align}
As can be easily checked, these two sets are obvious linear combinations of the two $t_{2g}$-derived $\Gamma_7$ states given in  Eqs. (\ref{two}) - (\ref{three}).

The level scheme of the crystal-field split states  is shown in Fig. \ref{Levels}. The Rashba spin-orbit splitting is possible when both spin-orbit interaction and broken inversion symmetry are present as indicated by the Rashba split parabolas for a $\Gamma_7$ state in  Fig. \ref{Levels}.   The strength of the Rashba splitting, characterized by the Rashba coefficient $\alpha_R$, depends on the proximity of the other crystal-field-split levels of the same symmetry, as the splitting comes from admixture of states of the same symmetry. Thus, for example, the $\Gamma_6$ states in Fig. \ref{Levels} should have  small Rashba coefficients because they are separated by the large crystal-field energy $\Delta$ from each other, allowing thereby only a small admixture between them. In fact, the Rashba coefficient for these states scales as $1/\Delta$ as seen from the Table I in the main text due to this energy denominator effect.
The only states that have a relatively larger $\alpha_R$ are the two $t_{2g}$-derived $\Gamma_7$ states owing to their proximity  in energy.

%: TB derivation
\section{Derivation of the Rashba Coefficients from the Tight-Binding model}

In order to understand the Rashba effect for the different Ta $d$ orbitals, we considered a tight-binding model for a cubic lattice terminated by a surface and subject to an electric field. The electron confinement in the direction normal to the surface due to the polar field and the surface potential leads to individual subbands, while the parallel  momentum  $k_{||}$ still remains a good quantum number. It is convenient to consider the Rashba effect for each subband, which may be described by taking the square lattice for the surface with the   ``atomic" orbitals being the appropriate Wannier functions for the subband. To this we add the spin-orbit interaction term ${\cal H}_{\rm SO}$ and the external electric field term 
${\cal H}_{\rm E}$, resulting in the minimal model Hamiltonian
\begin{equation}
{\cal H}={\cal H}_{\rm TB}+{\cal H}_{\rm SO}+{\cal H}_{\rm E}.
\label{eqH}
\end{equation}

Using the basis set: $z^2\uparrow$, $ z^2\downarrow$, $ x^2-y^2 \uparrow$, $ x^2-y^2 \downarrow$, $ xy\uparrow$, $xy\downarrow$,   $ xz\uparrow$, $ xz\downarrow$, $ yz\uparrow$, $ yz \downarrow $,
the tight-binding Hamiltonian in the momentum space is given by 
\begin{widetext}			
\begin{equation}
{\cal H}_{\rm TB} (\vec k) =
\begin{pmatrix} %z^2 & x^2-y^2 & xy & xz & yz \cr\hline
V_\sigma(c_x+c_y+4 c_z)/2+\Delta+\delta & 
-\sqrt{3}\,V_\sigma(c_x-c_y)/2 & 0 & 0 & 0\cr 
-\sqrt{3}\,V_\sigma(c_x-c_y)/2 & 
 3\,V_\sigma(c_x+c_y)/2+\Delta & 0 & 0 & 0\cr
 0 & 0 & 2\,V_\pi\,(c_x+c_y) +\epsilon & 0 & 0\cr 
0 & 0 & 0 & 2\,V_\pi\,(c_x+c_z)  & 0 \cr 
0 & 0 & 0 & 0 & 2\,V_\pi\,(c_y +c_z)  
\end{pmatrix}  
 \otimes
 \begin{pmatrix}
1 & 0 \cr
 0 & 1 \cr
\end{pmatrix}
\label{TB}
\end{equation}
where $c_x =\cos k_x$, $c_y =\cos k_y$, $c_z =\cos k_z$,
$V_\sigma$ and $V_\pi$ are the hopping integrals\cite{Harrison} between the $d$ orbitals and 
$\Delta$, $\varepsilon$, and $\delta$ are the crystal field parameters. The $2 \times 2$ matrix in the spin space in  the above equation is a unit matrix because the tight-binding hopping does not cause any  spin-flip scattering. We take $k_z = 0$ and the remaining energy shifts between 
the various $d$ orbitals due to surface confinement are incorporated into the crystal field parameters, viz., $\Delta, \delta,$ and $\varepsilon$, without any loss of generality.
The spin-orbit interaction part ${\cal H}_{\rm SO}=\xi \ {\vec L}\cdot {\vec S}$ in the same basis set, is given by
\begin{equation}
{\cal H}_{\rm SO} (\vec k)
=\frac{\xi}{2}\begin{pmatrix}0 & 0 & 0 & 0 & 0 & 0 & 0 & -\sqrt{3} & 0 & \sqrt{3}\,i\cr 0 & 0 & 0 & 0 & 0 & 0 & \sqrt{3} & 0 & \sqrt{3}\,i & 0\cr 0 & 0 & 0 & 0 & -2\,i & 0 & 0 & 1 & 0 & i\cr 0 & 0 & 0 & 0 & 0 & 2\,i & -1 & 0 & i & 0\cr 0 & 0 & 2\,i & 0 & 0 & 0 & 0 & -i & 0 & 1\cr 0 & 0 & 0 & -2\,i & 0 & 0 & -i & 0 & -1 & 0\cr 0 & \sqrt{3} & 0 & -1 & 0 & i & 0 & 0 & -i & 0\cr -\sqrt{3} & 0 & 1 & 0 & i & 0 & 0 & 0 & 0 & i\cr 0 & -\sqrt{3}\,i & 0 & -i & 0 & -1 & i & 0 & 0 & 0\cr -\sqrt{3}\,i & 0 & -i & 0 & 1 & 0 & 0 & -i & 0 & 0
\end{pmatrix},
\label{Eq-SO}
\end{equation}
where there is no momentum dependence because the spin-orbit interaction is a local term that couples the orbital and spin angular momenta on the same site only.

Finally, the electric field leads to non-zero matrix elements between several orbitals  due to the field-induced asymmetry of the orbital lobes,  as illustrated in Fig. 5 of the paper, and these matrix elements are expected to be roughly proportional to the local electric field.
We define these matrix elements as $\alpha=\langle xy|{\cal H}_E|xz\rangle_{\hat y} = \langle xy|{\cal H}_E|yz\rangle_{\hat x} $,
$\beta=\langle xz|{\cal H}_E|z^2\rangle_{\hat x}= \langle yz|{\cal H}_E|z^2\rangle_{\hat y}$, 
and $\gamma=\langle x^2-y^2|{\cal H}_E|yz\rangle_{\hat y}= \langle xz|{\cal H}_E|x^2-y^2\rangle_{\hat x}$, 
where the second orbital is located on the nearest-neighbor site along the direction indicated by  the subscript. Other matrix elements such as 
 $\langle yz|{\cal H}_E|z^2\rangle_{\hat x}= 0$  from symmetry.
 It is important to note that these matrix elements change sign if the hopping is from right to left or vice versa, which results in the sine factors in the Bloch sum (as opposed to a cosine factor which leads to the usual $k^2$ band dispersion in the tight-binding theory). These sine factors eventually show up as the linear-k terms in the Rashba Hamiltonian for small momentum. We thus have
\begin{equation}
{\cal H}_{\rm E} (\vec k) = 
 2i\begin{pmatrix}0 & 0 & 0 & -\beta\sin{k_x} & -\beta\sin{k_y}\cr
0 & 0 & 0 & -\gamma\sin{k_x} & \gamma\sin{k_y} \cr
0 & 0 & 0 & \alpha\sin{k_y} & \alpha\sin{k_x}\cr 
\beta\sin{k_x} & \gamma\sin{k_x} & -\alpha\sin{k_y} & 0 & 0\cr 
\beta\sin{k_y} & -\gamma\sin{k_y} & -\alpha\sin{k_x} & 0 & 0\cr 
\end{pmatrix}
 \otimes
 \begin{pmatrix}
1 & 0 \cr
 0 & 1 \cr
\end{pmatrix}.
\label{EqE} 
\end{equation}
\end{widetext}

%:Lowdin
{\it L\"owdin downfolding} -- 
The Hamiltonian Eq. \ref{eqH}  results in the spin-orbit split bands and we obtained the form of the band structure by downfolding the effects of other bands further away in energy via  the perturbative L\"owdin downfolding\cite{Lowdin}. 
The L\"owding downfolding is in essence a perturbative method, which works if the energies of the orbitals to be downfolded are far removed from the main orbitals of interest.
The L\"owdin downfolding procedure for solving the eigenvalue problem $(H - \lambda I) |\psi\rangle = 0$   is to partition the Hamiltonian into blocks
\begin{equation}
{\cal H}=\begin{pmatrix} h & b \cr b^\dagger & c\end{pmatrix}   , 
\end{equation}
where the two blocks are well separated in energy. We are not interested in the states in the block $c$, but include its effects on the block $h$ by perturbation theory. The exact result for the effective Hamiltonian for states in the $h$ subspace is given by
\begin{equation}
h^\prime = h + b ( \lambda I - c)^{-1} b^\dagger,
\label{downfolding}
\end{equation}
which however involves the eigenvalue $\lambda$ of the full Hamiltonian. It can be shown that an iterative solution of Eq. \ref{downfolding} produces the Brillouin-Wigner perturbation series,\cite{Ziman} which to the lowest order yields the result:
\begin{eqnarray}
h^\prime_{ij} = h_{ij} + \sum_k \frac{b_{ik} b_{kj}} {\lambda - c_{kk}},
\label{Heff}
\end{eqnarray}
where $i$ and $j$ belong to the subspace $h$ and $k$ belongs to the subspace $c$. In most cases $\lambda$ can be replaced by the diagonal elements $h_{ii}$ and Eq. \ref{Heff} is valid if  $| b_{ik} | \ll | h_{ii} - c_{kk} | $. 

%: Figure perturbation
\begin{figure} [t]
\scalebox{0.5}{\includegraphics{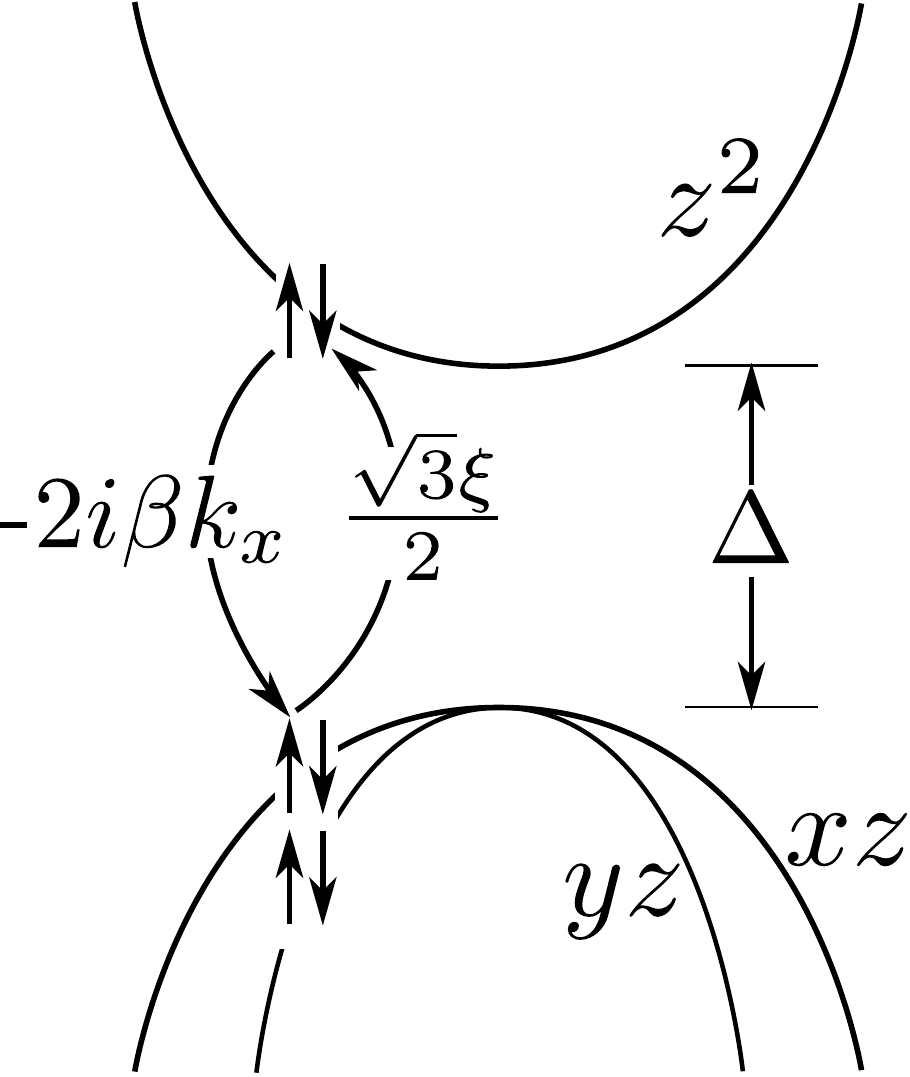}}
\caption{ Derivation of the Rashba coefficient for the $z^2$ orbital using the second-order perturbation theory. The figure clarifies 
how both the SOI, which produces the $\sqrt 3 \xi /2$ term, as well as the inversion asymmetry,
which produces the $-2 i \beta k_x$ term in the figure, are needed to couple the spin $\uparrow$ and $\downarrow$ states in the $z^2 $ band, leading to the Rashba spin splitting.
} 
\label{pert}
\end{figure} 

We can now use the L\"owding downfolding to derive an effective Hamiltonian for states that are well-separated in energy from the remaining states in the 
Hamiltonian Eq. (\ref{eqH}). We are interested in bands near the $\Gamma$ point ($\vec k =0$), where the bands are quadratic in momentum without the Rashba effect. We therefore retain only the linear terms in momentum in the final expression for the downfolded Hamiltonian.
For example, by downfolding the effect of the remaining eight orbitals in the full Hamiltonian into the ($z^2 \uparrow, z^2 \downarrow$) subspace via Eq. \ref{Heff} with 
 $\lambda=3V_\sigma+\Delta+\delta$, we get the result, for small $k$ around the $\Gamma$ point:
\begin{equation}
{\cal H}_{z^2}=\begin{pmatrix}3\,V_\sigma+\Delta+\delta & \frac{2\,\sqrt{3}\,\beta\,\xi\,\left(k_y+i\,k_x\right) }{4\,V_\pi-3\,V_\sigma-\Delta-\delta}\cr 
\frac{2\,\sqrt{3}\,\beta\,\xi\,\left(k_y-i\,k_x\right) }{4\,V_\pi-3\,V_\sigma-\Delta-\delta} & 3\,V_\sigma+\Delta+\delta\end{pmatrix}.  
\label{DF}
\end{equation}
In terms of the Pauli matrices, this is written as
\begin{equation}
{\cal H}_{z^2}= (3V_\sigma+\Delta+\delta) -\frac{2\sqrt{3}\beta\xi}{\Delta}(k_y\sigma_x-k_x\sigma_y),
\label{R}
\end{equation}
where in writing the second term, we have ignored the smaller energies $V_\sigma, V_\pi, \delta \ll \Delta$ in the denominator of the off-diagonal matrix elements in the downfolded Hamiltonian Eq. \ref{DF}. 
The second term in Eq. \ref{R} describes the momentum-dependent spin splitting and can be written in the familiar  Rashba form, viz.,
${\cal H}_R = \alpha_R (\vec k \times  \vec \sigma)    \cdot \hat z$,
with the Rashba coefficient $\alpha_R = - 2\sqrt{3}\beta\xi /  \Delta$ as given in the first line of Table I in the main text.  
The Rashba coefficients for the other orbitals were obtained in a similar manner.

{\it Second-order perturbation theory} -- It is instructive to illustrate the origin of the Rashba Hamiltonian using the second-order Brillouin-Wigner perturbation theory Eq. (\ref{Heff}). We illustrate this for the Rashba spin splitting of the $z^2$ orbital. 
Referring to Fig. \ref{pert}, the tight-binding band structure is indicated by the parabolas, which are spin-degenerate.

The spin degeneracy is removed due to the interaction via the intermediate orbitals $xz$ or $yz$.
 The various matrix elements can be read off from the Hamiltonians, Eqs. (\ref{TB} - \ref{EqE}), and two of them have been shown in Fig. \ref{pert}. Thus the off-diagonal matrix element $h'_{\uparrow\downarrow} \equiv \langle z^2 \uparrow| h' |z^2 \downarrow \rangle$ due to the interaction with the $|xz \uparrow \rangle$ state is given by the product of the two interactions indicated in Fig. \ref{pert} divided by the energy denominator, viz., 
$ (-2i \beta k_x) (\sqrt 3 \xi /2)/ \Delta$. Note that the first interaction is due to the electric-field induced asymmetry, while the second term is due to the spin-orbit interaction. Clearly, both are necessary to produce a non-zero result.
Adding up the contributions from the three other intermediate orbitals, viz., $|xz \downarrow \rangle$, $|yz \uparrow \rangle$, and $|yz \downarrow \rangle$,   we  get the desired result $h'_{\uparrow\downarrow} = - 2\sqrt 3 \beta \xi   (k_y + i k_x) /\Delta$ and, in the same manner, we can obtain the remaining matrix elements to yield
\begin{equation}
h' = 
\begin{pmatrix}  h'_{\uparrow\uparrow}  & h'_{\uparrow\downarrow} \cr 
 h'_{\downarrow\uparrow} & h'_{\downarrow\downarrow}
\end{pmatrix}  
= 
\begin{pmatrix}  \Delta  & \alpha_R(k_y+i k_x)\cr 
 \alpha_R(k_y-i k_x) & \Delta
\end{pmatrix}, 
\end{equation}
where the Rashba coefficient $ \alpha_R = -2 \sqrt 3\beta \xi / \Delta$ as before.

%:References

\end{document}